\documentclass[prl,aps,a4paper,10pt, twocolumn,superscriptaddress,preprintnumbers,a4paper,amsmath,amssymb,showpacs,floatfix]{revtex4}
\usepackage{graphicx,subfigure}
\usepackage{color}\color[rgb]{0.000,0.000,0.000} %used for font color
\usepackage{amssymb} %maths
\usepackage{amsmath} %maths
\usepackage[normalem]{ulem} %emphasize weiterhin kursiv
\usepackage {ulem} %\emph{Text}: Text wird unterstrichen

\begin{document}

\newcommand{\tfo}{TbFeO$_3$\ }
\newcommand{\Tc}{$T_\mathrm{C}$}
\newcommand{\Tn}{$T_\mathrm{N}$}
\newcommand{\Tntb}{$T_\mathrm{N}$(Tb)}
\newcommand{\Tnfe}{$T_\mathrm{N}$(Fe)}
\newcommand{\Fe}{Fe$^{3+}$}
\newcommand{\Tb}{Tb$^{3+}$}
\newcommand{\hc}{H$\| c$}

\newcommand{\Mueff}{$\mu_{\mathrm{eff}}$}
\newcommand{\MuB}{$\mu_\mathrm{B}$}

%\preprint{}

\title{Solitonic lattice and Yukawa forces in the rare earth orthoferrite \tfo}

\author{Sergey Artyukhin}
\author{Maxim Mostovoy}
\email[Email of corresponding author: ]{m.mostovoy@rug.nl}
\affiliation{Zernike Institute for Advanced Materials, University of Groningen, Nijenborgh 4, 9747 AG, Groningen, The Netherlands}
\author{Niels Paduraru Jensen}
\affiliation{Ris\o\ National Laboratory for Sustainable Energy,
Frederiksborgvej 399, DK-4000 Roskilde, Denmark}
\author{Duc Le}
\affiliation{Helmholtz-Zentrum Berlin f\"ur Materialien und Energie, Hahn-Meitner Platz 1, D-14109, Berlin, Germany}
\author{Karel Prokes}
\affiliation{Helmholtz-Zentrum Berlin f\"ur Materialien und Energie, Hahn-Meitner Platz 1, D-14109, Berlin, Germany}
\author{Vin\'{\i}cus G. Paula}
\affiliation{Helmholtz-Zentrum Berlin f\"ur Materialien und Energie, Hahn-Meitner Platz 1, D-14109, Berlin, Germany}
\affiliation{Instituto de F\'{\i}sica, Universidade do Estado do Rio de Janeiro, Rio de Janeiro, RJ 20559-900, Brazil}
\author{Heloisa N. Bordallo}
\affiliation{Helmholtz-Zentrum Berlin f\"ur Materialien und Energie, Hahn-Meitner Platz 1, D-14109, Berlin, Germany}
\affiliation{Nanoscience Center, Niels Bohr Institute, University of Copenhagen, DK-2100 Copenhagen, Denmark}
\affiliation{European Spallation Source ESS AB, P.O Box 176, SE-221 00 Lund, Sweden}
\author{Andrey Maljuk}
\affiliation{Helmholtz-Zentrum Berlin f\"ur Materialien und Energie, Hahn-Meitner Platz 1, D-14109, Berlin, Germany}
\affiliation{Leibniz Institute for Solid State and 
Materials Research Dresden, Helmholtzstra�e 20,
01069 Dresden, Germany}
\author{Sven Landsgesell}
\affiliation{Helmholtz-Zentrum Berlin f\"ur Materialien und Energie, Hahn-Meitner Platz 1, D-14109, Berlin, Germany}
\author{Hanjo Ryll}
\affiliation{Helmholtz-Zentrum Berlin f\"ur Materialien und Energie, Hahn-Meitner Platz 1, D-14109, Berlin, Germany}
\author{Bastian Klemke}
\affiliation{Helmholtz-Zentrum Berlin f\"ur Materialien und Energie, Hahn-Meitner Platz 1, D-14109, Berlin, Germany}
\author{Sebastian Paeckel}
\affiliation{Helmholtz-Zentrum Berlin f\"ur Materialien und Energie, Hahn-Meitner Platz 1, D-14109, Berlin, Germany}
\author{Klaus Kiefer}
\affiliation{Helmholtz-Zentrum Berlin f\"ur Materialien und Energie, Hahn-Meitner Platz 1, D-14109, Berlin, Germany}
\author{Kim Lefmann}
\affiliation{Nanoscience Center, Niels Bohr Institute, University of Copenhagen, DK-2100 Copenhagen, Denmark}
\author{Luise Theil Kuhn}
\affiliation{Ris\o\ National Laboratory for Sustainable Energy,
Frederiksborgvej 399, DK-4000 Roskilde, Denmark}
\affiliation{European Spallation Source ESS AB, P.O Box 176, SE-221 00 Lund, Sweden}
\author{Dimitri N. Argyriou}
\email[Email of corresponding author: ]{dimitri.argyriou@esss.se}
\affiliation{Helmholtz-Zentrum Berlin f\"ur Materialien und Energie, Hahn-Meitner Platz 1, D-14109, Berlin, Germany}
\affiliation{European Spallation Source ESS AB, P.O Box 176, SE-221 00 Lund, Sweden}

\date{\today}
%\preprint{}

%\pacs{ENTER PACS HERE}
\begin{abstract}
The control of domains in ferroic devices lies at the heart of their potential for technological applications.  Multiferroic materials offer another level of complexity  as domains can be either or both of a ferroelectric and magnetic nature. Here we report the discovery of a novel magnetic state in the orthoferrite TbFeO$_{3}$ using neutron diffraction under an applied magnetic field. This state has a very long incommensurate period ranging from $340$ \AA\ at 3K to $2700$ \AA\ at the lowest temperatures and exhibits an anomalously large number of  higher-order harmonics, allowing us to identify it with the periodic array of sharp domain walls of Tb spins separated by many lattice constants. The Tb domain walls interact by exchanging spin waves propagating through the Fe magnetic sublattice. The resulting Yukawa-like force, familiar from particle physics, has a finite range that determines the period of the incommensurate state.
\end{abstract}

\maketitle

Materials with magnetic transition metal and rare earth ions show a variety of spectacular effects originating from the coupling between the two spin subsystems. The transition metal spins interact stronger and order at higher temperatures than the spins of rare earth ions, but they are also much less anisotropic. That is why their orientation can be controlled by the rare earth magnetism. Such re-orientation transitions observed in many rare earth ferrites, chromites and manganites have profound effects on their magnetic, optical and elastic properties   \cite{Belov_Uspekhi_1976,Buchelnikov_Uspekhi_1996,Kimel_Nature_2005}.

\begin{figure}[htb!] \includegraphics[height=\columnwidth,angle=270]{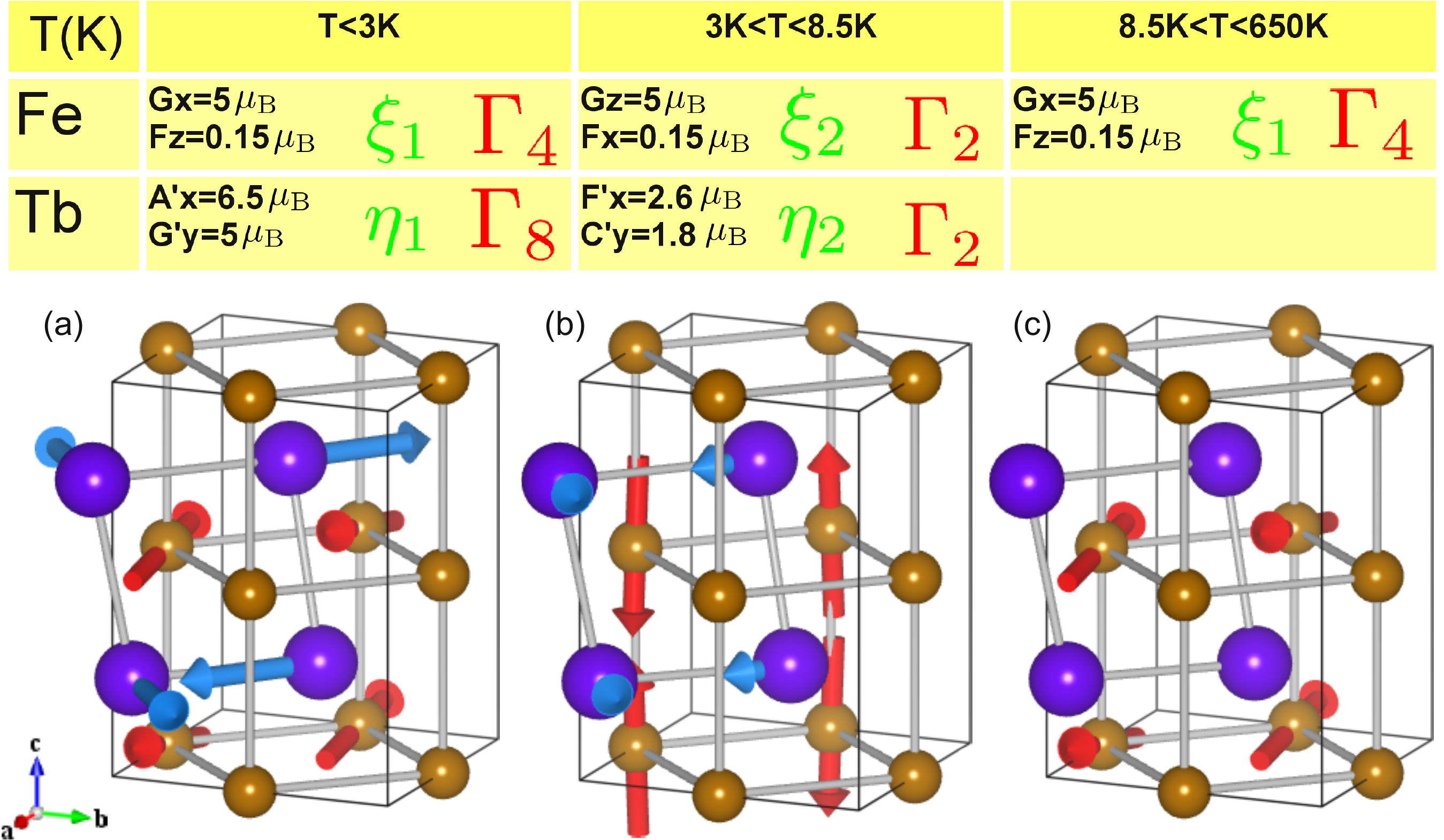}
\scriptsize
\caption{Magnetic ordering of Fe$^{3+}$ (brown spheres) and Tb$^{3+}$ (blue spheres) ions in the three uniform phases of TbFeO$_3$: the LT phase (panel a), the IT phase (panel b) and the HT phase (panel c). Also shown are the corresponding order parameters, irreducible representations and experimental values of magnetic moments \cite{Bertaut_1967}. The various types of magnetic order depicted here are labeled as  
 F for a ferromagnetic ordering, G for a the two-sublattice 
antiferromagnetic N\'eel state, A for ferromagnetic $ab$ planes stacked antiferromagnetically 
and finally  C for ferromagnetic chains parallel to the $c$ axis coupled antiferromagnetically.}
 \label{fig:magOrder}
\end{figure}

Recently it was realized that interactions between transition metal and rare earth spins also play an important role in multiferroic and magnetoelectric materials \cite{Kimura_Nature_2003,Hur_Nature_2004,Zvezdin_JETPLetters_2008,Tokunaga_NatMat2009} .
Thus the coupling between the Mn spins forming a spiral state in the multiferroic TbMnO$_3$ and the Ising-like Tb spins leads to a significant enhancement of the electric polarization induced by the spiral \cite{Prokhnenko_PRL_2007,Aliouane_JPCM_2008}. In GdFeO$_3$  orthoferrite the polarization only appears when   the independent magnetic orders of Fe and Gd sublattices are present simultaneously \cite{Tokunaga_NatMat2009}, while in DyFeO$_3$ the interplay between the spins of Fe and Dy ions gives rise to one of the strongest linear magnetoelectric responses observed in single-phase materials \cite{Tokunaga_PRL_2008}. 

\tfo\ is an orthorhombic perovskite (space group P$bnm$) where Fe spins order antiferromagnetically in what is called G-type order along the $a$ axis and ferromagnetically (F-type) along the $c$ axis as shown on Fig.~\ref{fig:magOrder}c.  This type of commensurate spin order, denoted as $G_{x}F_{z}$, has an onset at approximately \Tnfe$=650$K. On cooling in zero field \tfo\ undergoes two transitions driven by Tb-Tb and  Tb-Fe interactions \cite{Bertaut_1967,Bouree_1975}. The ordering of Tb spins at \Tntb$\sim$8.5K occurs simultaneously with a rotation of Fe spins in the $ac$ plane, so that below 8.5K ferromagnetic components of both Fe and Tb spins align along the $a$ axis, while their antiferromagnetic components are orthogonal to each other.  The magnetic configuration of this intermediate temperature (IT) phase is $F_{x}G_{z}$ for Fe, and $F'_{x}C'_{y}$ for Tb (see Fig.~\ref{fig:magOrder}b). However, below $\sim$ 3K there is an additional spin re-orientation transition to a low temperature (LT) phase which flips the Fe spins back to their higher temperature $G_{x}F_{z}$ order, while the Tb spins order antiferromagnetically in the  $A'_{x}G'_{y}$ state (see Fig.~\ref{fig:magOrder}a).  

Single crystals of \tfo\ were grown under 4 bar oxygen pressure using the crucible-free floating zone method. Their quality was checked by X-ray diffraction. Neutron diffraction experiments were made on a large single crystal of TbFeO$_{3}$, at the BER-II reactor of the Helmholtz Zentrum Berlin using the FLEX cold triple-axis spectrometer with collimation of 60$'$-60$'$-60$'$, $k_{i}$=1.3\AA$^{-1}$, and a cooled Be filter positioned in the scattered beam. Additional measurements were made also with the E4 two-axis diffractometer with $\lambda$=2.8\AA. In both cases a magnetic field was applied along the $c-$axis of the sample using a superconducting horizontal field magnet. Dielectric measurements were performed at the Laboratory for Magnetic Measurements at the Helmholtz-Zentrum Berlin (LaMMB), with temperatures varying between 0.3~K and 15~K and with magnetic fields up to 1.9~T. Magnetic Field and temperature control was provided by an Oxford Instruments 14.5~T cryomagnet equipped with a Heliox $^3$He~insert. An Andeen-Hagerling 2700A Capacitance Bridge was used to measure the capacitance and loss of a disc shaped sample of TbFeO$_3$, which was mounted between the electrodes of a parallel-plate capacitor.

\begin{figure}[tb!]
\centering
\includegraphics[scale=0.44]{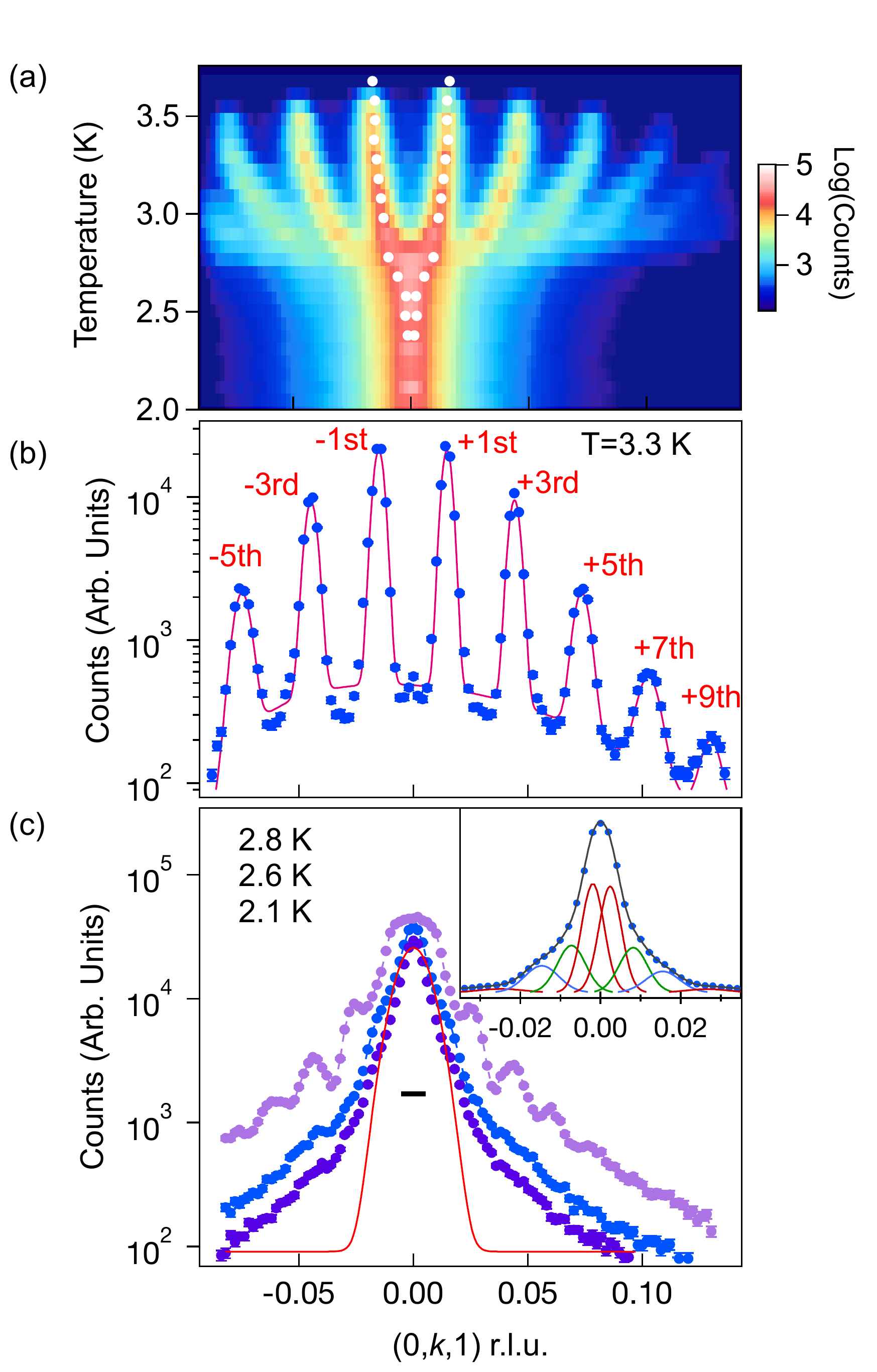}
\scriptsize
\caption{Single crystal neutron diffraction data measured on cooling and in a magnetic field parallel to the $c$ axis of \hc = 2T. All scans are measured in reciprocal space along (0,$k$,1). In all panels except for the insert in panel (c), the data is plotted on a logarithmic scale in order to show the weaker higher harmonic reflections. (a) Temperature-dependent neutron diffraction measurements are represented in a two-dimensional plot with intensity depicted as color on a log scale shown on the right of the panel. White circles are positions of the first harmonic reflection computed by fitting the diffraction data at each temperature to gaussian peak profiles. (b) The scan measured at 3.3K is plotted here while the various harmonic reflections are labeled accordingly. The data are shown as circles while the continuous line is a fit of a series of Gaussian profiles to the data. (c) Scans measured at 2.8, 2.6 and 2.1 K show here the transition from the IC phase as the $\epsilon$ rapidly decreases to a smaller values into the LT-IC phase.  At 2.8K the various higher harmonic reflections are clearly still evident, while at lower temperatures they merge closer together and appear almost as a single peak.  For the 2.1K measurement we show the fit of a single gaussian peak to the data which fails to account for the  observed peak profile. The instrumental resolution is depicted here as a horizontal bar.  In the inset we show a fit to the same 2.1K data to a series of gaussians, up to the 7th harmonic,  that provide a better model to the data.  The IC wavenumber obtained from this fit is $\epsilon$=0.002(1).}
\label{fig:harmonics}
\end{figure}

Using single crystal neutron diffraction we have probed the A,C,G and F-type orders in TbFeO$_{3}$ by tracking the intensity of the corresponding magnetic Bragg reflections in zero field and in an applied field along the $c$ axis (see Methods for experimental details). In zero magnetic field our results are consistent with the previously observed  sequence of the re-orientation and inverse re-orientation transitions \cite{Bertaut_1967}. Above $\sim$8.5K we find only G-type reflections, while the development of ferromagnetic order is evident from the enhanced intensities of lattice Bragg reflections. Below 8.5K we find G- and C-type reflections, while below 3K only A- and G-type reflections can be discerned.

In an applied magnetic field (\hc) we find a far more complex behavior.  Here we performed a series of field cooled measurements, while monitoring accessible A- and G-type reflections.
In Fig.~\ref{fig:harmonics}(a) we show in the form of a color plot the temperature dependence of scans along $k$ around the A-type (001) reflection.  At high temperatures this reflection is absent as there is no order of an A-type component for either the Fe or Tb magnetic sublattice as indicated in Fig. 1.  However, on cooling a series of reflections appears below 3.8K that seem to merge into a single peak below $\sim$2.8K.  Examination of the wavevector of these reflections easily establishes that they are odd harmonics of up to 11th order. The wavevector of the 1st harmonic is $\mathbf{Q}=(0, \epsilon,1)$ with $\epsilon\sim$ 0.015 r.l.u. [see Fig.~\ref{fig:harmonics}(b)].  The incommensurate periodicity of this phase that we shall refer to as IC, is approximately 67 units cells or $\sim$340 \AA. The full width at half maximum (FWHM) of these reflections is relatively sharp giving a coherence length of $\sim$700 \AA\ or approximately two full cycles of this unusual order.

The physical significance of these observations is that the Tb-spin order in a \hc\ field develops a square-wave modulation -- a periodic array of widely separated domain walls in Tb magnetic order. We ascribe the observed scattering to be dominated by Tb spins first because of its substantially greater intensity compared to the higher temperature Fe order found around for example G-type reflections and also because in zero field A-type reflections are associated only with Tb spin order. In the modulated A-state, Tb spins form ferromagnetic stripes in the $ab$ planes with the width 170 \AA\ along the $b$ axis. The $a$ component of magnetization alternates from stripe to stripe, while the stacking of spins along the $c$ axis is antiferromagnetic. Investigation of an Fe G-type reflection under the same condition suggests that Fe-spins are weakly perturbed by this unusual Tb-order (see supplementary information).

\begin{figure}[b!]
\centering
\includegraphics[width=\columnwidth]{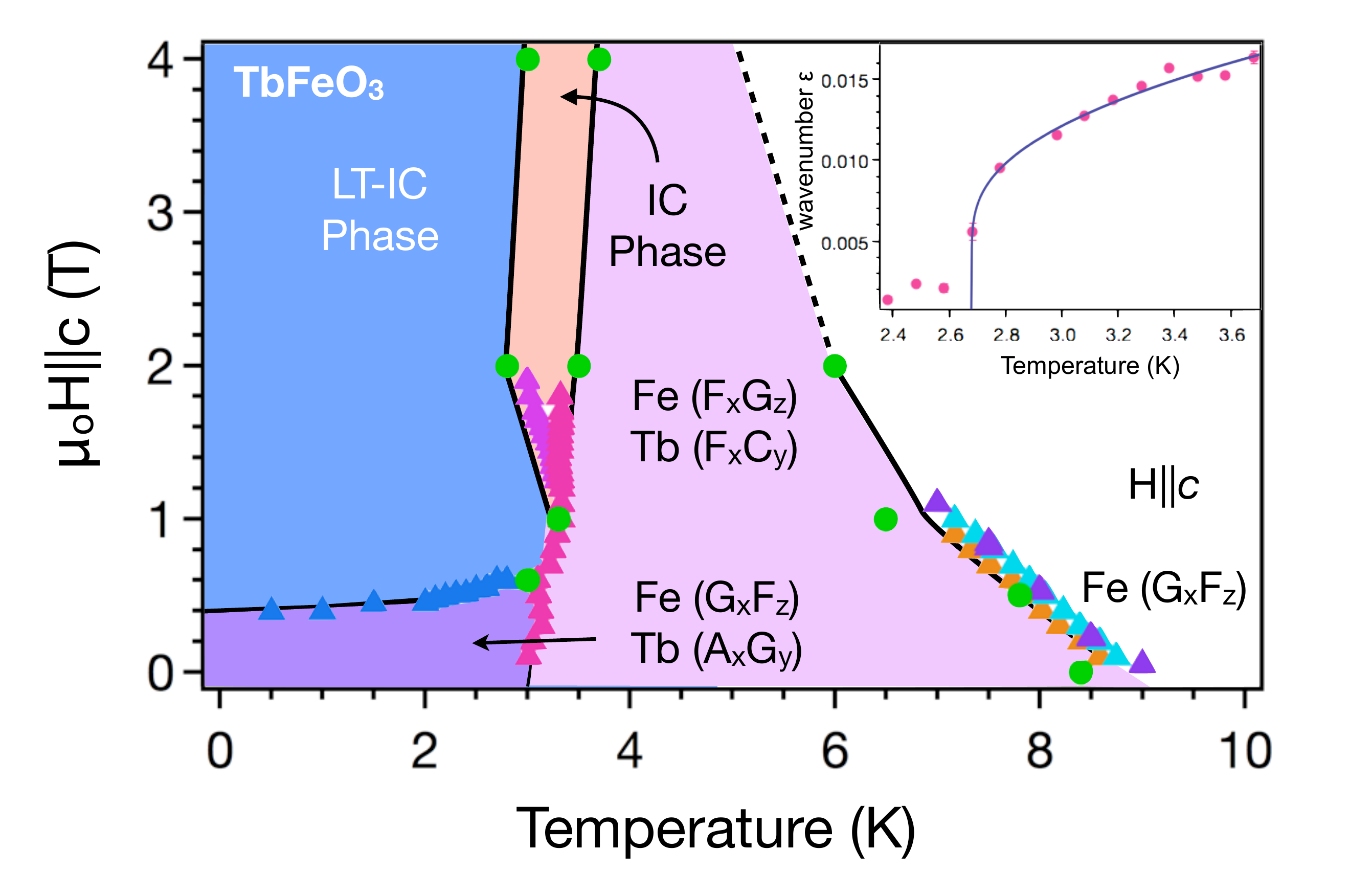}
\scriptsize
\caption{The magnetic phase diagram of \tfo\ determined from neutron diffraction data (shown as circles) and capacitance and loss measurements (shown as triangles) that are described in the supplementary information (see Fig. S2-S5). In the insert we show the temperature dependence of the modulation wavevector $\epsilon$ measured in an applied magnetic field $\mu_0H_c = 2$T. The blue line is the fit obtained using the theoretical  description of the IC state in terms of the periodic kink array.}
\label{fig:expphasediag}
\end{figure}

On cooling below 2.8K Fig.~\ref{fig:harmonics}(a) would indicate that the Tb modulation abruptly disappears and the Tb subsystem returns to the zero field state with the uniform $A'_{x}G'_{y}$ order. However, closer inspection of the diffraction data indicates that the (001) reflection on cooling does not yield a simple gaussian peak shape. Rather as shown in Fig.~\ref{fig:harmonics}(c), there is  clear evidence of more than one peak at 2.8K and significant deviations from a gaussian peak shape at 2.1 K.  This indicates that below  this inverse re-orientation transition the state remains incommensurate with a much larger periodicity than above the transition.  Using carefully constrained fits of Gaussians to the data measured below 3.0K (see insert of Fig.~\ref{fig:harmonics}(c)) it is clear that with lowering temperature $\epsilon$ decreases sharply below 2.8K but does not go to zero to yield a pure commensurate (001) reflection. Rather from the temperature  dependence of $\epsilon$, shown in the inset of Fig.~\ref{fig:expphasediag}, it is evident that below 3.7K, $\epsilon$ decreases gradually with cooling until 2.8K, below which it drops abruptly to a value of $\sim$0.002.  This low temperature incommensurate (LT-IC) state has  a periodicity of approximately 500 units cells or 2700 \AA, which distinguishes it from the higher temperature IC phase with a smaller  periodicity.

We have performed measurements similar to those shown in Fig.~\ref{fig:harmonics}(a) at various fields as well as two zero field cooled measurements at 3.0 and 3.3 K where the field was subsequently applied isothermally in order to map out the various transitions that occur in this lower temperature regime in TbFeO$_{3}$. In addition, we have conducted capacitance and loss measurements between 0.3 to 10 K in a magnetic field (\hc) between 0 and 1.9T.  The transitions that are evident in the neutron data are  correlated with anomalies in both the capacitance and loss data (see supplementary information), which allowed us to construct the phase diagram shown in Fig.~\ref{fig:expphasediag}. Interestingly, the capacitance and loss data imply that if the sample is cooled in zero field below $\sim$3~K and then field is applied, the transition into the LT-IC phase is not reversible and this state can then be stabilized in zero field (see Fig. S5). 

%\noindent{\it Theory}:
Next we discuss the nature of interactions stabilizing such an unusual periodic domain wall array and holding the domain walls at large distances from each other. Well below \Tnfe$\sim650$K  the magnitude of the ordered antiferromagnetic moment of the Fe subsystem is independent of temperature, while its direction in the $ac$ plane described by the angle $\theta$ can significantly vary due to the low magnetic anisotropy of the \Fe\ ions. In our notations $\xi_1 = \cos \theta$ is the order parameter of the $G_x$ state, while $\xi_2 = \sin \theta$ describes the $G_z$ ordering. The free energy density of the Fe subsystem is
\begin{equation}\label{eq:fFe}
f_{\rm Fe} = \frac{c}{2} \left(\frac{d\theta}{dy}\right)^2 + \frac{K}{2}\sin ^2\theta-h \cos \theta,
\end{equation}
where the first term describes the exchange between Fe spins along the $b$ axis, the second term is the magnetic anisotropy, which for $K > 0$ favors the $G_x$ order, and the last term is the Zeeman interaction with the magnetic field $H_z$ in the $G_xF_z$ state.

The free energy of Tb spins is expanded in powers of the order parameters $\eta_1$,  describing the zero phase LT state with antiparallel Tb spins in neighboring $ab$ layers (Fig.~\ref{fig:magOrder}a), and $\eta_2$, describing the IT state with parallel Tb spins in neighboring layers (Fig.~\ref{fig:magOrder}b):
\begin{eqnarray}\label{eq:fTb}
f_{\rm Tb}= &~& \frac{c_1}{2}\left(\frac{d\eta_1}{dy}\right)^2+\frac{c_2}{2}\left(\frac{d\eta_2}{dy}\right)^2
+\frac{a_1}{2}\eta_1^2+\frac{a_2}{2}\eta_2^2 \nonumber \\
&~&+ \frac{b_1}{4}\eta_1^4+\frac{b_{12}}{2}\eta_1^2\eta_2^2 + \frac{b_2}{4}\eta_2^4+\ldots.
\end{eqnarray}
For $\Delta = a_2 - a_1 > 0$ the Tb subsystem would order in the state with $\eta_1 \neq 0$ below some temperature $T_0$, at which $a_1 = 0$. However, the interaction between the Tb and Fe spins favors the IT state with $\eta_{2} \neq 0$ and $\theta = \pm \frac{\pi}{2}$, in which both subsystems have a ferromagnetic moment along the $a$ axis. Since $\eta_2$ and $\xi_2 = \sin\theta$ transform in the same way (see the supplemental material), this interaction is a linear coupling,
\begin{equation}\label{eq:fint}
f_{\rm Fe-Tb}= -\lambda \xi_{2} \eta_{2}.
\end{equation}
For $\lambda^2>\Delta K$, the `unnatural' IT state with parallel Tb spins in neighboring layers and Fe spins rotated by $90^{\circ}$ away from the easy axis, intervenes between the states with the `natural' orders of Fe and Tb spins. In this way one obtains the zero-field phase diagram of TbFeO$_3$\cite{Bertaut_1967,Belov_JETP_1979}.

\begin{figure}[b!]
 \centering
 \includegraphics[width=0.7\columnwidth]{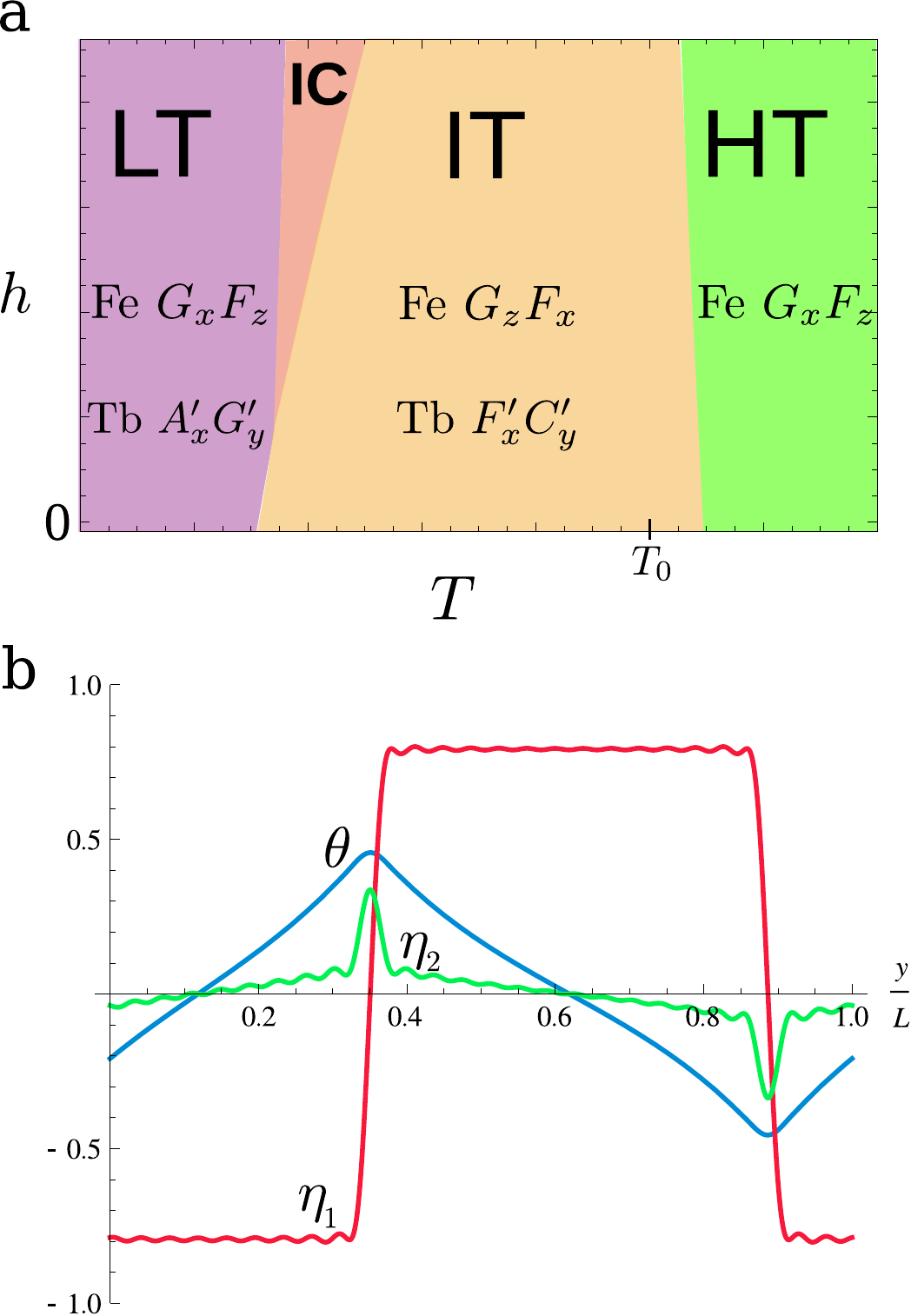}
 \scriptsize

 \caption{(a) Magnetic phase diagram of the Landau model of \tfo\  including the Fe-Tb interaction described by the Lifshitz invariants Eq.(\ref{eq:fL2}). The parameters used to obtain this phase diagram are: $\Delta = 0.5$, $K = 0.125$, $\lambda = 0.275$, $b_1 = b_2 = 1.0$, $b_{12} = 2.4$, $g_1 = 0.187$, $g_2 = 0$, $c = 1$, and $c_1 = c_2 = 0.01$. (b) The $y$-dependence of the Tb order parameters, $\eta_1$ (red line) and $\eta_2$ (green line), and the angle $\theta$ measured in radians (blue line) describing the rotation of Fe spins in the IC state with the period $L = 340$\AA.}
 \label{fig:phasediag}
\end{figure}

In addition, we consider the so-called Lifshitz invariants linear in order parameter gradients,
\begin{equation}\label{eq:fL2}
f_{\rm L} = g_1 \left(\eta_{1} \partial_{y} \xi_2 - \xi_2 \partial_y \eta_{1}\right)+g_2 \left(\eta_{1} \partial_{y} \eta_{2} - \eta_{2} \partial_y \eta_{1}\right),
\end{equation} 
which favor the experimentally observed  periodic spin modulation along the $b$ axis. Similar terms   inducing modulations along the $a$ and $c$ axes are forbidden by symmetry (see the supplemental material). Minimizing the total free energy -- the sum of Eqs.(\ref{eq:fFe}-\ref{eq:fL2}) -- we obtain the phase diagram shown in Fig.~\ref{fig:phasediag}a, which includes a narrow incommensurate (IC) phase region, which we assign to the same IC phase revealed in our neutron data. 

It is important to stress the difference between the IC state in TbFeO$_3$ and the long period spin spirals in non-centrosymmetric magnets, also described  using Lifshitz invariants \cite{Dzyaloshinskii_JETP_1964}.   First, the crystal lattice of \tfo\ is centrosymmetric (inversion symmetry is only broken in the LT phase by the Tb spin ordering). Equation (\ref{eq:fL2}) is the interaction between two distinct magnetic phases: the LT Tb state (odd under inversion) and the IT phase (even under inversion). It is only effective close to the boundary where these two phases have equal free energies, which is why the IC state is observed in a very narrow region of the phase diagram. 

Second, spirals in non-centrosymmetric magnets result from the relatively  weak spin-orbit coupling \cite{Dzyaloshinskii_JETP_1964,Moriya_PR_1960}. On the other hand, the coupling Eq.(\ref{eq:fL2}) likely originates from a stronger Heisenberg exchange: in the supplemental material we give symmetry arguments showing that the exchange interactions between the Tb and Fe spin orders varying along the $b$ axis do not cancel. Furthermore, the coupling between two Tb order parameters [the second term in Eq.(\ref{eq:fL2})] resulting from  interactions between rare earth spins separated by relatively long distances, is expected to be much weaker than interactions between the Tb and Fe spins described by the first term (in our calculations $g_2 = 0$).

Third and most important, the observation of the large number of Fourier harmonics in the IC state of  TbFeO$_3$ shows that this state is qualitatively different from a magnetic spiral with slowly varying spin vectors. To account for the difference between the isotropic Fe spins and the Ising-like Tb spins \cite{Bouree_1975,Bidaux75}, we assumed that $c_1,c_2 \ll c$ and allowed for 25 harmonics in the  periodic modulation of order parameters when we minimized the free energy. The resulting incommensurate state is shown in Fig.~\ref{fig:phasediag}b. While the angle $\theta$ describing the Fe spins undergoes small amplitude fan-like oscillations around zero, corresponding to the oscillations of the weak ferromagnetic moment of Fe ions around the applied magnetic field $H \|c$, the low-temperature Tb order parameter $\eta_1$ exhibits sudden jumps.

\begin{figure}[h]
\centering
\includegraphics[width=\columnwidth]{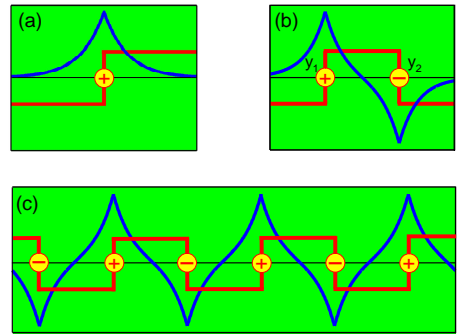}
\scriptsize

\caption{(a) Domain wall (kink) in Tb ordering (red line) and the angle $\theta$ (blue line) describing the perturbation of Fe spins near the domain wall; (b) kink-antikink pair; (c) periodic array of domain walls with alternating charges.}
\label{fig:electrostatics}
\end{figure}

To understand the nature of the force that holds these atomically sharp domain walls at  distances of $\sim 170$ \AA\ from each other, we (briefly) discuss an interesting field-theoretical interpretation of the coupled  system of rare earth and transition metal spins. Consider a single domain wall located at $y = 0$ where the Ising-like LT order parameter $\eta_1$ shows a discontinuous jump from  $-|\eta_1|$ to $+|\eta_1|$ or vice versa (see Fig.~\ref{fig:electrostatics}a). Such a kink can be assigned the topological charge $Q = (\eta_1(+\infty) - \eta_1(-\infty))/2|\eta_1| = \pm 1$. The free energy per unit area of the domain wall is the `bare' energy $F_{DW}^{(0)}$ resulting from interactions between Tb spins plus
\begin{equation}\label{eq:Ftheta}
F_{\theta} = -2g\theta(0)Q+
\frac{1}{2}\int\!\!dy
\left[c\left(\frac{d\theta}{dy}\right)^2 + \left(K+h\right)\theta^2\right],
\end{equation}
where the first term is the Lifshitz invariant Eq.(\ref{eq:fL2}) ($g = g_1|\eta_1|$ and $g_2 = 0$), describing the interaction between the Tb and Fe spins, while the second term is the free energy of Fe spins for $|\theta| \ll 1$. Equation (\ref{eq:Ftheta}) can be interpreted as an energy of a charged plane with the surface charge density $gQ$ (a `nucleon') interacting  with the field $\theta$, which describes spin waves in the Fe magnetic subsystem playing the role of pions with the mass $m = \sqrt{\frac{K+h}{c}}$.  Minimizing $F_{\theta}$ with respect to $\theta(y)$, we obtain the distortion in the Fe spin ordering produced by the Tb domain wall, $\theta(y) = \frac{Qg}{\sqrt{c(K+h)}}e^{-\frac{|y|}{l}}$ (see Fig.~\ref{fig:electrostatics}a), which reduces the domain wall free energy:
\begin{equation}
F_{DW} = F_{DW}^{(0)} - \frac{g^2}{\sqrt{c(K+h)}}.
\end{equation}   

When $F_{DW}$ becomes negative, the domain walls tend to condense. Their density is, however, limited by Yukawa-like interactions between the domain walls, which result from the exchange of massive particles and have the range $l = m^{-1} =  \sqrt{\frac{c}{K+h}}$. These interactions attract equal `electric' charges and repel  opposite ones. Neighboring domain walls in a periodic array have opposite `electric' charges, since topological charges of the domain walls alternate along the $b$ axis. The interaction  between two neighboring domain walls located at $y_1$ and $y_2$  [see Fig.~\ref{fig:electrostatics}(b)] is
\begin{equation}
U(y_2-y_1) = \frac{g^2}{\sqrt{c(K+h)}} e^{-\frac{|y_2-y_1|}{l}},
\end{equation} 
and the total `electrostatic' free energy of an array of domain walls with the charges $\{Q_n\}$ alternating along the $b$ axis (including the `self-energy' of the charged surfaces) is given by 
\begin{equation}
F_{\theta} = - \sum_{n,m}Q_n U(y_n - y_m) Q_m,
\end{equation}
where $y_n$ is the position of the $n$-th kink. Minimizing the free energy density for an equidistant array of kinks (see Fig.~\ref{fig:electrostatics}(c)), we obtain the optimal period of the incommensurate state. Its temperature dependence fits well the experimental data above 2.8K, as shown in the inset of Fig.~\ref{fig:expphasediag}. The length scale for the period of the IC state, set by $l\sim150$\AA, is essentially the thickness of the domain wall in the antiferromagnetic ordering of Fe spins, even though such walls are not present in the IC state. Thus the long period of the IC state of Tb spins originates from the large stiffness and low magnetic anisotropy of the Fe magnetic subsystem.  

The nature of the LT-IC phase remains a puzzle, as the low temperature state of our theoretical model, in which the domain wall free energy becomes positive, is uniform. One possibility is that the IC order at the lowest temperatures is short-ranged. Such a state might be formed by isolated metastable domain walls stabilized by impurities and crystal imperfections and separated by the average distance of $\sim$1500 \AA.  Such a conclusion can not be excluded from the experimental data due to the small value of $\epsilon$ for this state. 

The tantalizing suggestion from our work is that periodic domain wall arrays may be present in other orthoferrites and orthochromites, since they have similar zero field magnetic phases. Therefore, identifying domain states as the one we describe here is important for understanding their magnetoelectric and multiferroic properties. Domain walls induced by interactions between the rare earth and transition metal magnetic subsystems may also be relevant for the applications of this class of materials, as the dynamics of these domain walls, coupled by long-ranged interactions can strongly affect the switching of the spontaneous electric polarization with an applied magnetic field and \textit{vice versa}.\\

\noindent{\it Acknowledgements: } DNA and SL thank the Deutsche Forschung Gemeinschaft for support under contract AR 613/2-1. The neutron scattering work was supported by the Danish National Research Council through DANSCATT. NPJ thanks Prof. J�rn Bindslev Hansen, for support and guidance.

%%%%%%%%%%%%%%%%%%%%%%%
\clearpage
\begin{center}
\textbf{Supplemental material}
\end{center}

\section{Symmetry analysis}

We use the so-called Bertaut's notations\cite{Bertautnotation} to describe the symmetry of magnetic states of TbFeO$_3$ (see Table~\ref{tab:orders}). The ordering of Fe spins in the HT phase is described by the $\Gamma_4 $ representation. In the IT state both Fe and Tb spin orders have $\Gamma_2$ symmetry, while in the LT phase the Fe order changes back to $\Gamma_4$, while the ordering of Tb spins has $\Gamma_8$ symmetry (see Fig.~\ref{fig:magOrder}). Furthermore, the Lifshitz invariant has the form 
\begin{equation}
\Gamma_8 \partial_y \Gamma_2 - \Gamma_2 \partial_y \Gamma_8,
\end{equation}
from which Eq.(\ref{eq:fL2}) follows. We note that the rotationally symmetric scalar products,
\begin{equation}
\mathbf{A}' \cdot \partial_y \mathbf{F} - \mathbf{F} \cdot \partial_y \mathbf{A}' \;\;\;\;\mbox{and}\;\;\;\; 
\mathbf{G}' \cdot \partial_y \mathbf{C} - \mathbf{C} \cdot \partial_y \mathbf{G}',
\end{equation}
where e.g. $\mathbf{A}' \cdot \partial_y \mathbf{F} = A'_x \cdot \partial_y F_x+A'_y \cdot \partial_y F_y+A'_z \cdot \partial_y F_z $, are also invariant under all transformations of the P$bnm$ group showing that the coupling between inhomogeneous rare earth and transition metal magnetic orders  can originate from Heisenberg exchange interactions\cite{Aliouane_JPCM_2008}.

\begin{table}[htbp]
\centering
\begin{tabular}{|c|c|c|c|c|c|}
\hline
& Fe & Tb & ${\tilde m}_{x}$ &  ${\tilde m}_{y}$ & $m_{z}$\\
[0.4ex]
\hline
$\Gamma_1 $&$A_x G_y C_z$& $C'_{z}$& $+$ & $+$ & $+$\\
[0.4ex]
$\Gamma_2  $&$F_x C_y G_z$& $F'_{x} C'_{y}$& $+$ & $-$ & $-$\\
[0.4ex]
$\Gamma_3  $&$C_x F_y A_z$& $C'_{x} F'_{y}$& $-$ & $+$ & $-$\\
[0.4ex]
$\Gamma_4 $&$G_x A_y F_z$& $F'_{z}$& $-$ & $-$ & $+$\\
[0.4ex]
$\Gamma_5  $&& $G'_{x} A'_{y}$& $-$ & $-$ & $-$\\
[0.4ex]
$\Gamma_6  $&& $A'_{z}$& $-$ & $+$ & $+$\\
[0.4ex]
$\Gamma_7  $&& $G'_{z}$& $+$ & $-$ & $+$\\
[0.4ex]
$\Gamma_8  $&& $A'_{x} G'_{y}$& $+$ & $+$ & $-$\\
[0.4ex]
\hline
\end{tabular}
\caption{Transformation properties of representations of P$bnm$ space group under the three generators of the group: the two glide mirrors, ${\tilde m}_x: (x,y,z) \rightarrow (1/2-x,1/2+y,z)$ and ${\tilde m}_y: (x,y,z) \rightarrow (1/2+x,1/2-y,1/2+z)$, and the mirror $m_z: (x,y,z) \rightarrow (x,y,1/2-z)$.}
\label{tab:orders}
\end{table}

\section{Measurements of the (011) G-type reflection }

Although the magnetic environment in our diffraction experiment constrains the portions of reciprocal space that we can access, we were able to also probe a limited region around the G-type reflection (0,1,1). In a field \hc = 2T we observe IC reflections below 3.5K with the same incommensurability $\epsilon$ and temperature dependence as for the satellites around the A-type (0,0,1) reflection (see Fig. S1 in the supplementary information). 
The commensurate (0,1,1) G-type reflection observed above 3.5K arises only from the Fe-spin ordering (see Fig. 1). The fact that the intensity of the commensurate reflection does not vary through this transition suggests that the Fe order is not significantly perturbed, in agreement with our theory.

\begin{figure}[ht]
%\centering
\begin{center}
\includegraphics[width=\columnwidth]{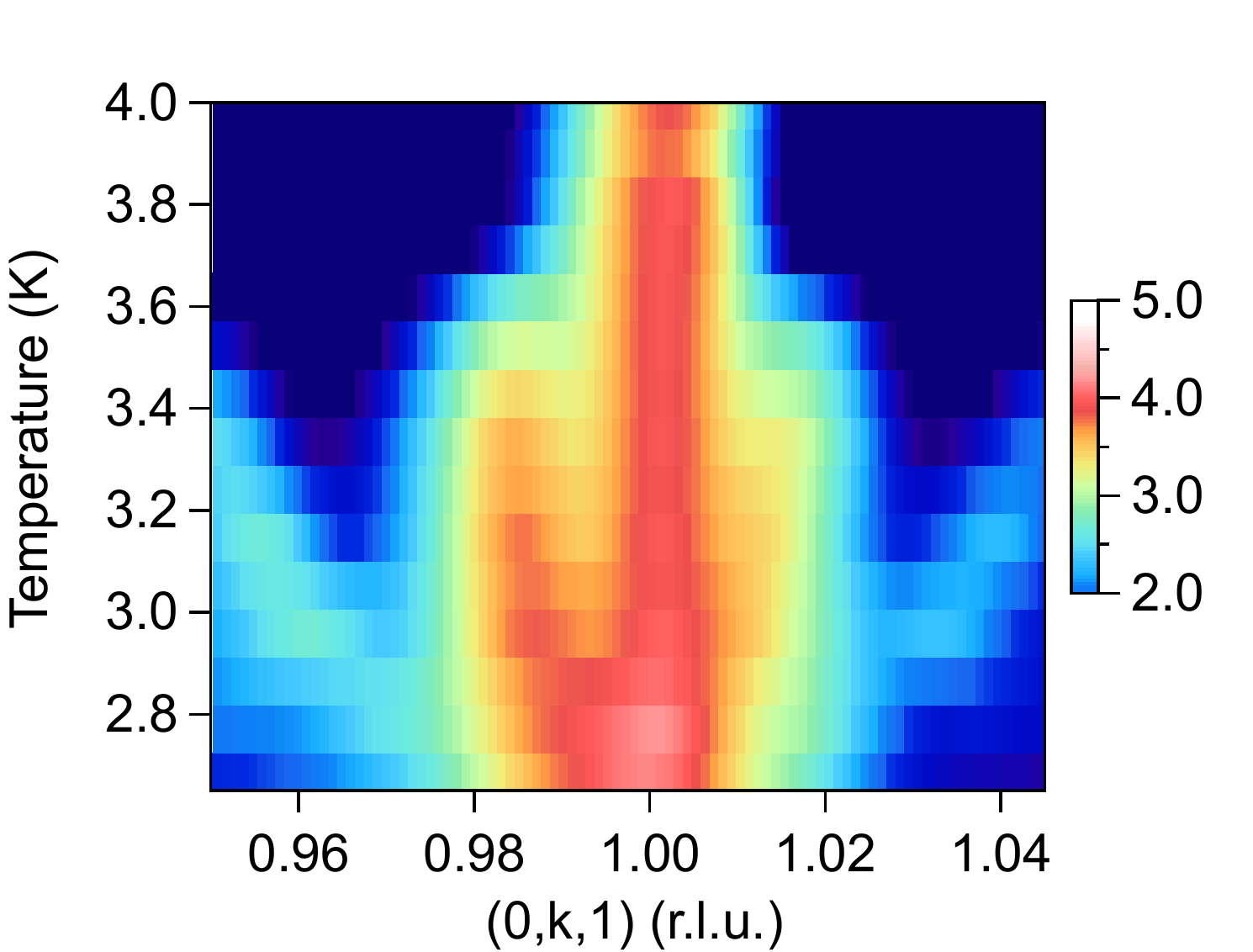}
\end{center}
\raggedright
FIG. S1: Single crystal neutron diffraction data measured on cooling and in a magnetic field parallel to the $c$-axis of \hc=2T. Scans are measured in reciprocal space along (0,$k$,1) around the G-type reflection (0,1,1). As above the temperature dependent neutron diffraction measurements are represented in a two-dimensional plot with intensity depicted as color on a log scale shown on the right of the panel. The 1st and 3rd harmonic reflections are evident below 3.4 K. 
\label{fig:Gtype}
\end{figure}

%\clearpage

\section{Dielectric Measurements}

Measurements of the capacitance and loss were performed both as a function of temperature between 0.3 and 15K in field cooled mode and isothermally as a function of magnetic field up to 1.9~T, in zero field cooled mode. In our measurements,  a calibrated Cernox resistance thermometer (CX-1030) on the sample holder was used to monitor the sample temperature, which was measured with an LakeShore 370 AC Resistance Bridge. The resolution of the temperature measurement is $5\cdot 10^{-4}$, although the temperature stability decreases around 3~K, due to the boiling temperature of liquid $^3$He of 3.2~K. The capacitance and loss measurements were performed at a frequency of 1000~Hz and an excitation of 15~V. The resolution of the capacitance measurement is $2\cdot 10^{-5}$.

\begin{figure}[htbp]
	\centering
  	\begin{minipage}[t]{7.2 cm}	       
      \includegraphics[width=1\textwidth]{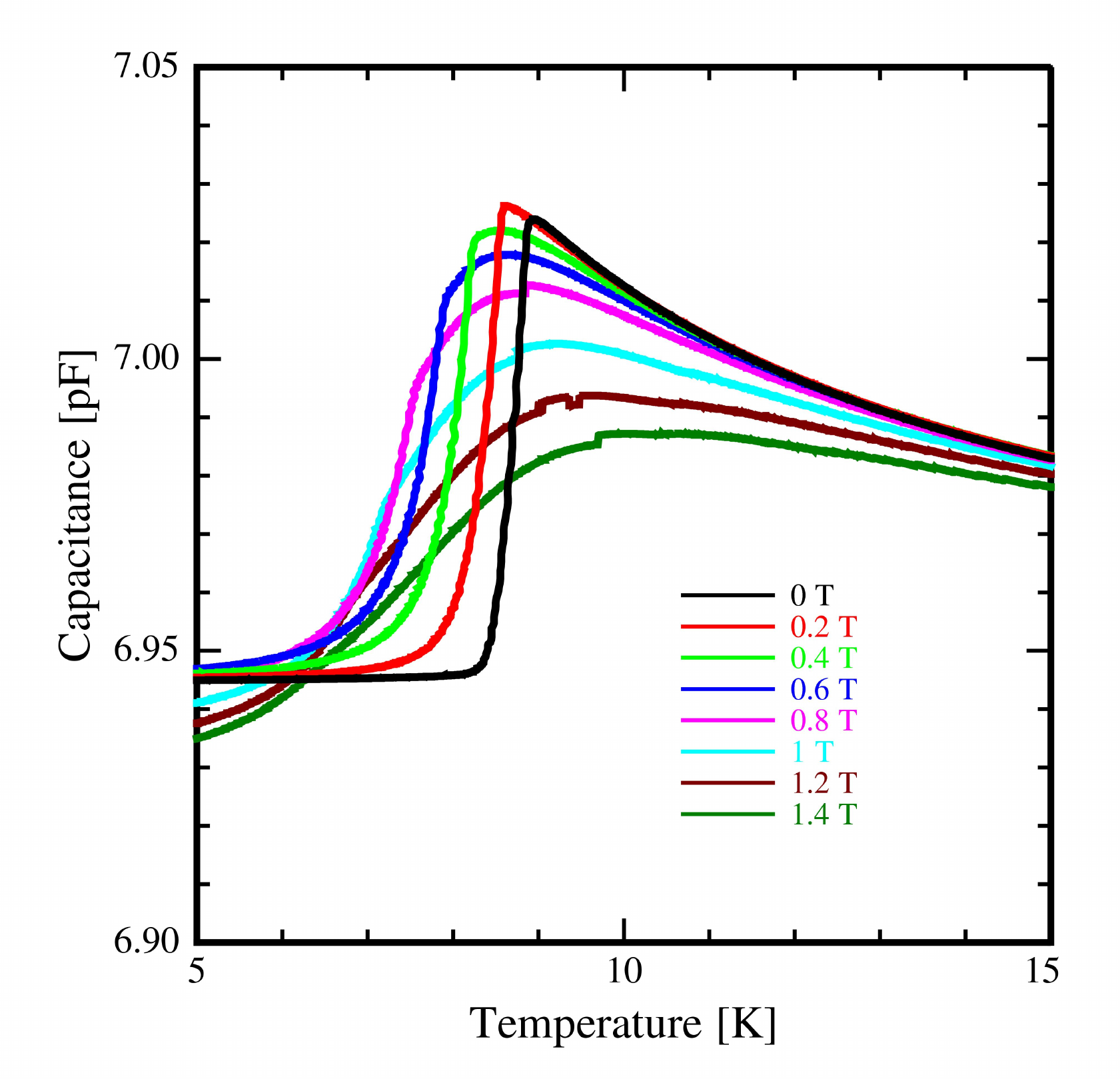}
   \end{minipage}
   \begin{minipage}[t]{7.2 cm}
     \includegraphics[width=1\textwidth]{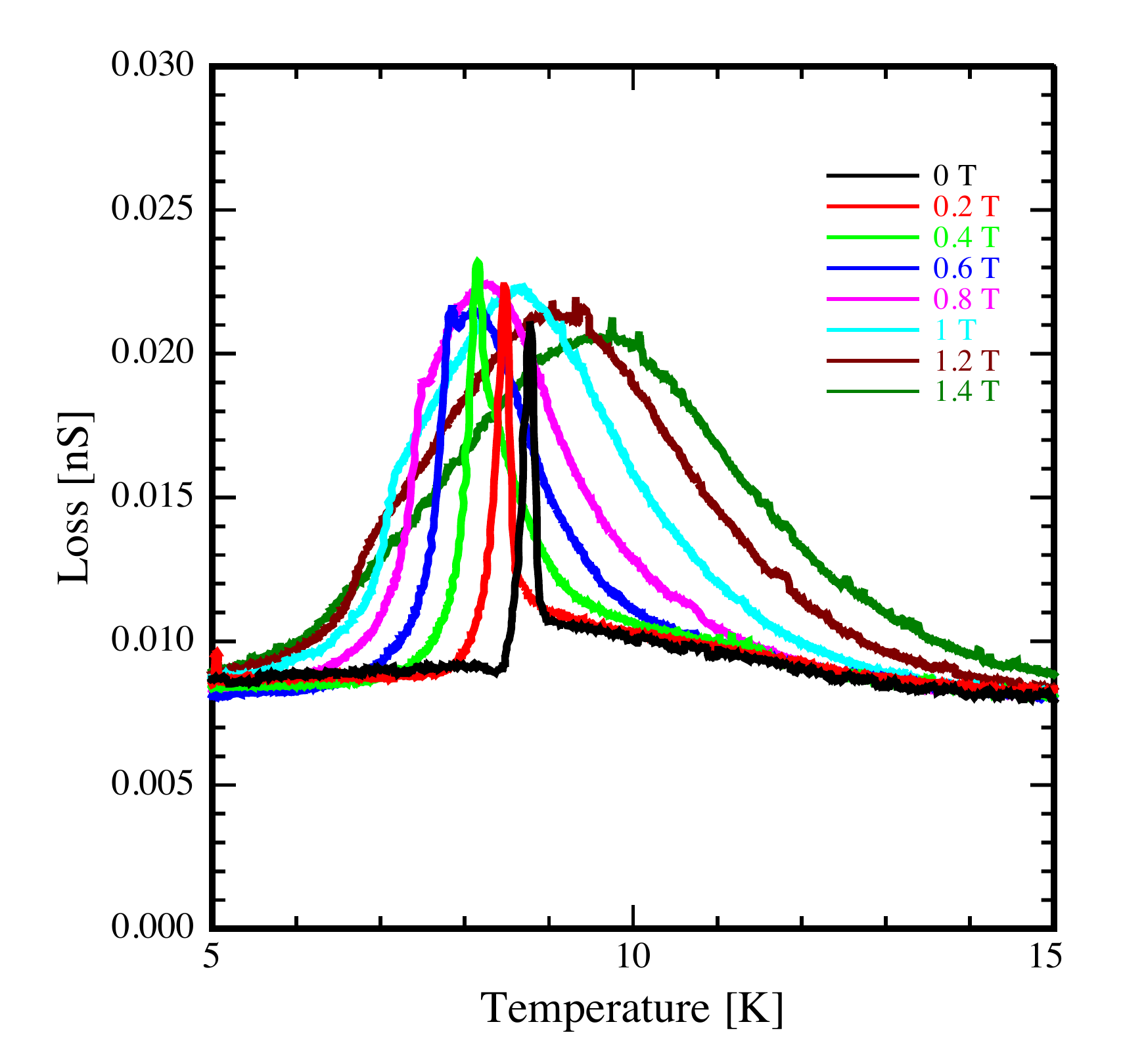}
	\end{minipage}
	\label{fig:TScanHigh}
	
	\raggedright
FIG. S2: Example data of capacitance and loss measured from a single crystal of TbFeO$_{3}$ between 0 and 1.5~T as a function of temperature between 5 and 15~K. Here the sample was cooled with an applied magnetic field and measurements were taken continuously during warmup with a sweep rate of 125~mK per minute. We found that measurement on cooling showed exactly the same behavior. We also noted that that measurements after field cooling and zero field cooling do not differ in this temperature range.
\end{figure}

\begin{figure}[htbp]
	\centering
  	\begin{minipage}[t]{7.2 cm}	       
      \includegraphics[width=1\textwidth]{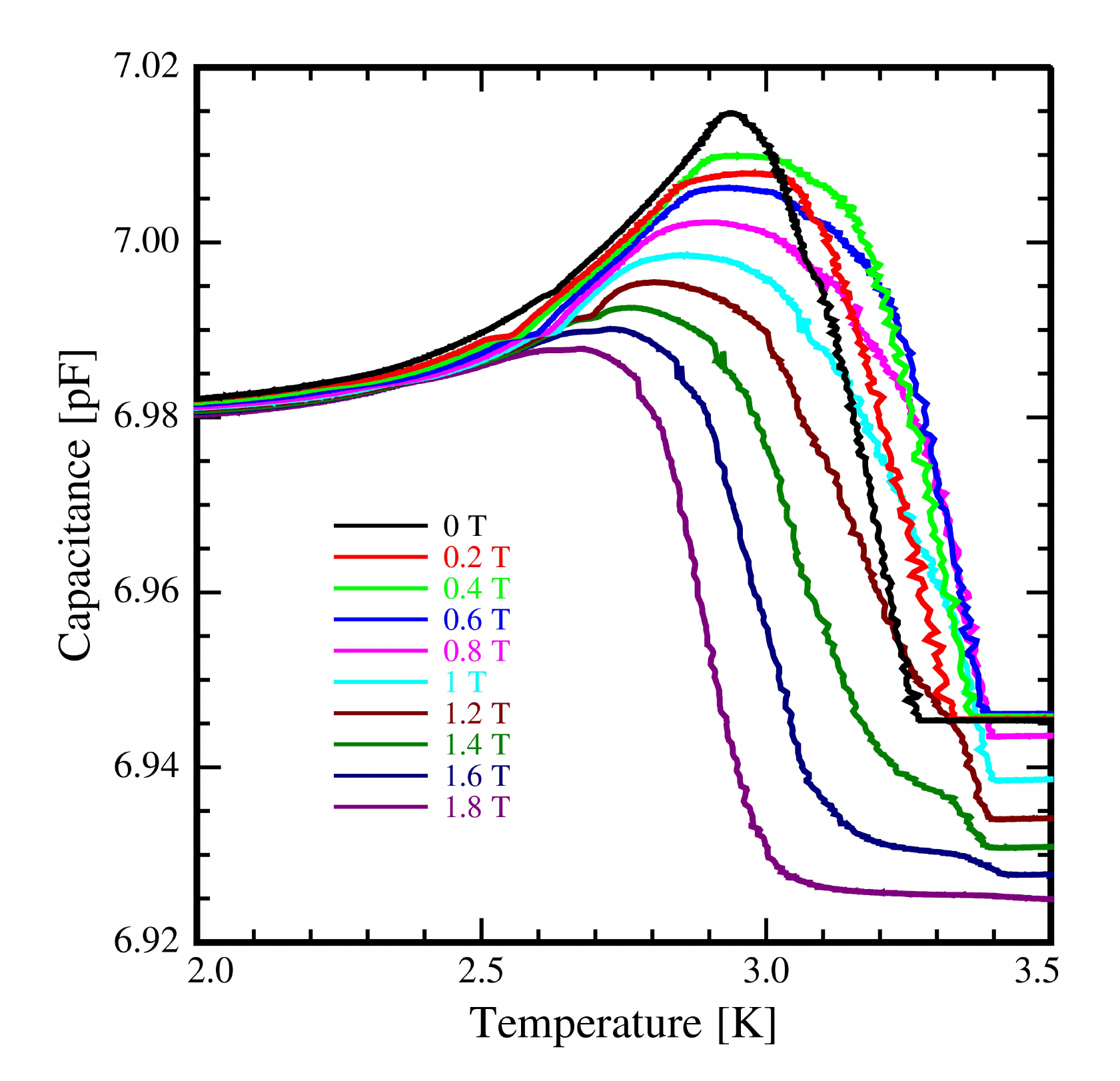}
   \end{minipage}
   \begin{minipage}[t]{7.2 cm}
     \includegraphics[width=1\textwidth]{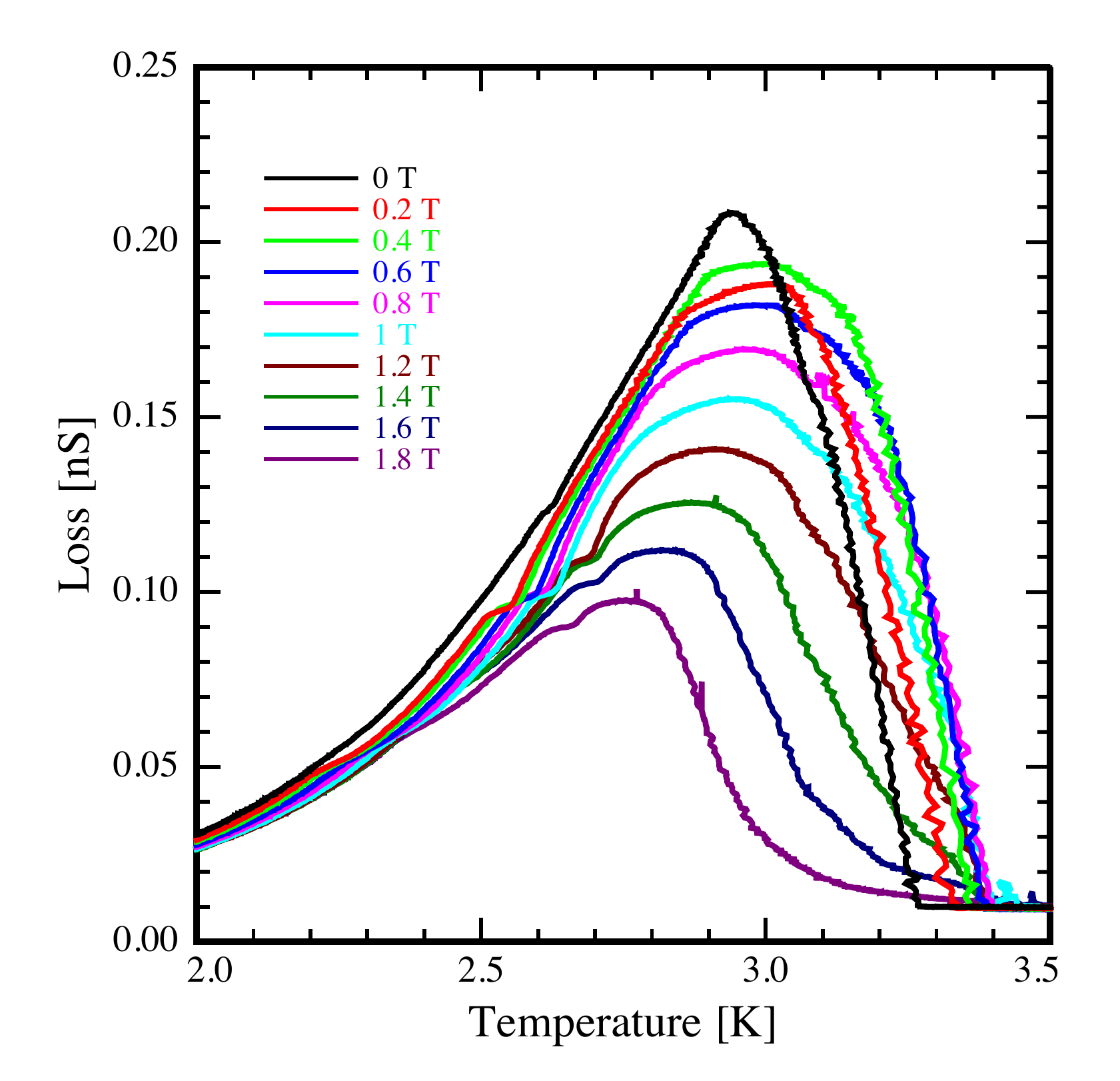}
	\end{minipage}
  \label{fig:TScanLow}
\raggedright

FIG. S3: Example data of capacitance and loss measured from a single crystal of TbFeO$_{3}$ between 0 and 1.9~T as a function of temperature between 2 and 3.5~K. Here the sample was cooled with an applied magnetic field while measurements were taken continuously during warming with a sweep rate of 30~mK per minute. Measurements were also performed down to 0.3~K, but no anomalies were found indicating the absence of further transitions. 
	
\end{figure}

\begin{figure}[htbp]
	\centering
  	\begin{minipage}[t]{7.2 cm}	       
      \includegraphics[width=1\textwidth]{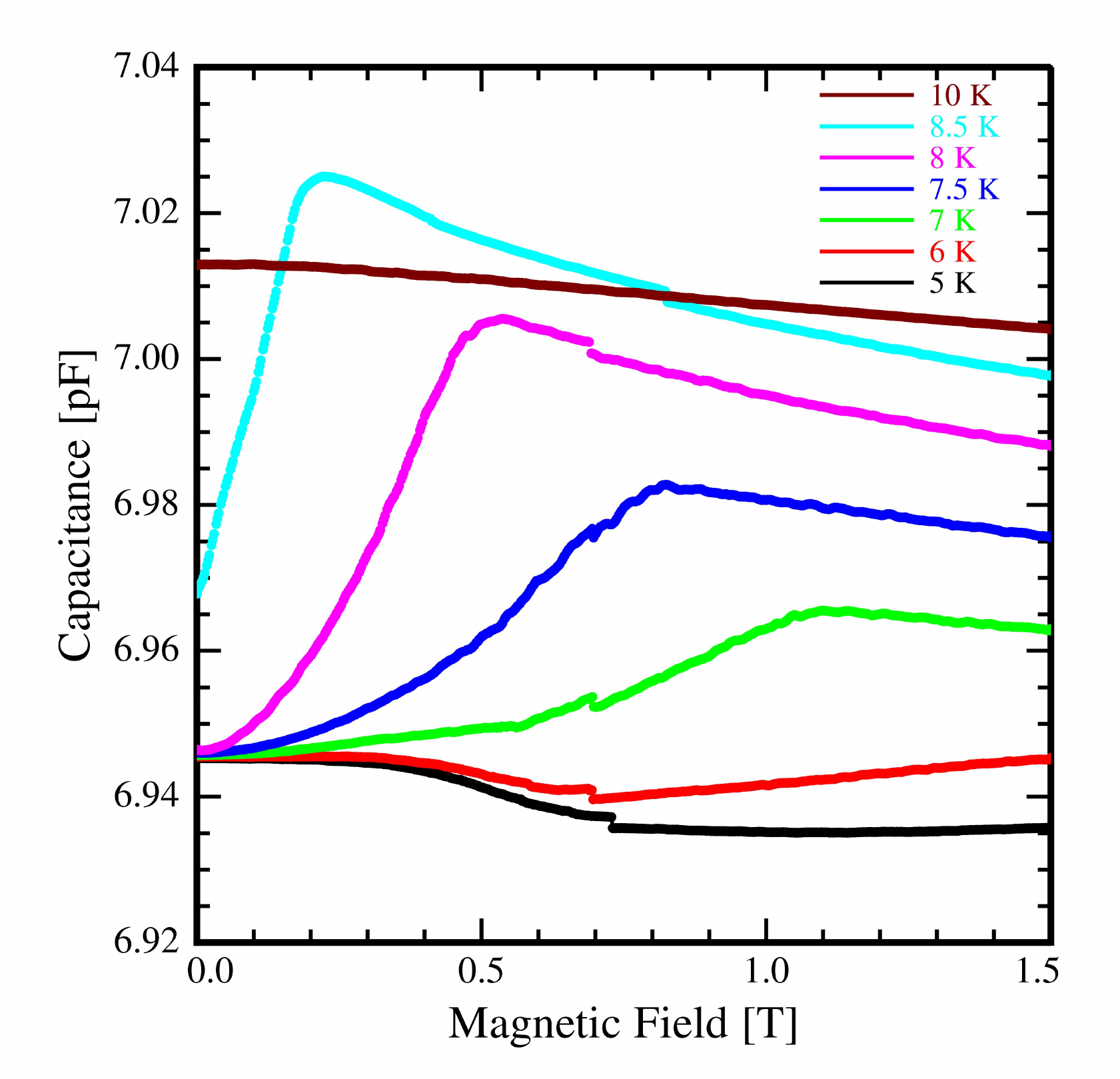}
   \end{minipage}
   \begin{minipage}[t]{7.2 cm}
     \includegraphics[width=1\textwidth]{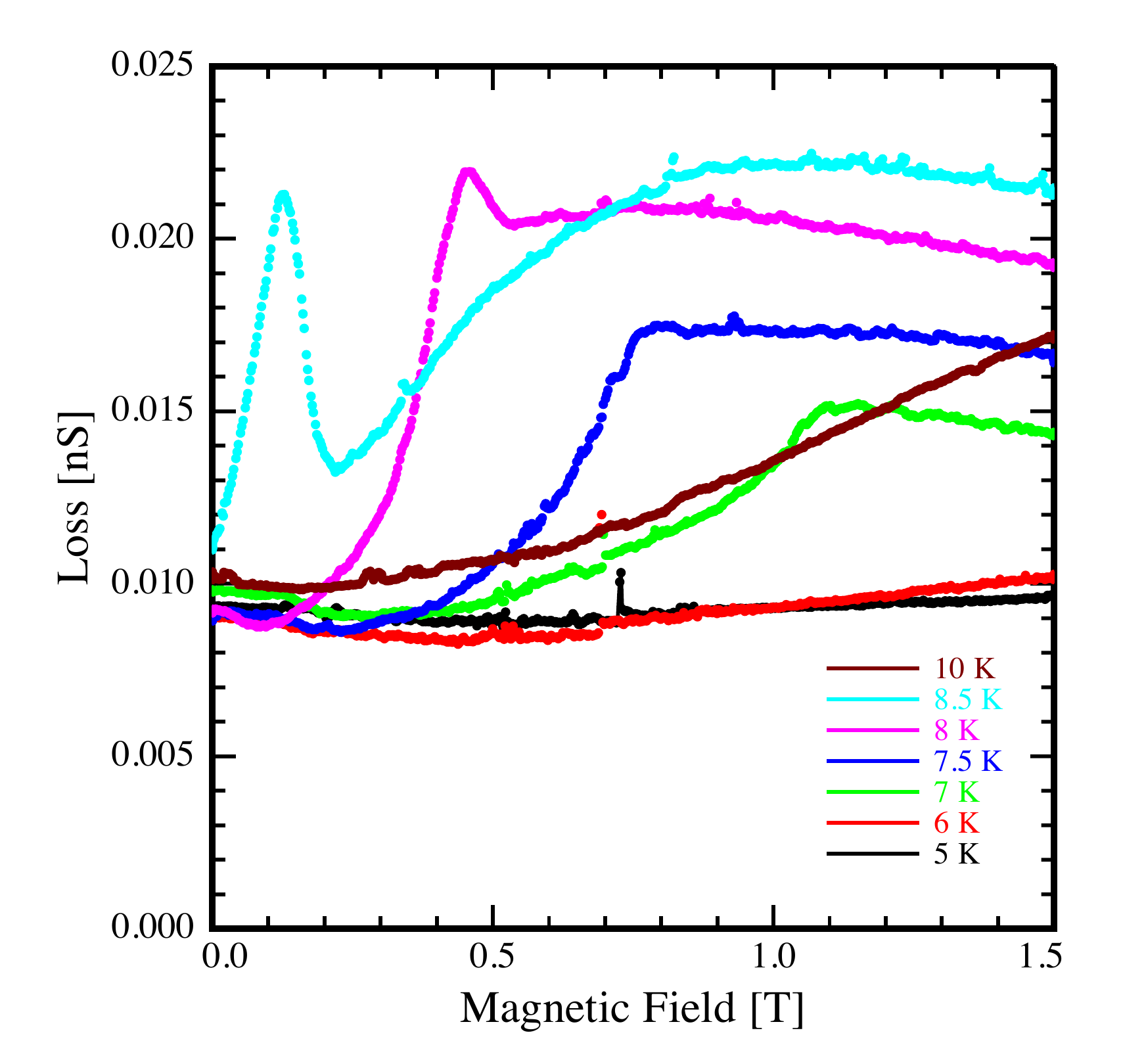}
	\end{minipage}
  \label{fig:BScanHigh}
\raggedright

FIG. S4: Example data of capacitance and loss measured from a single crystal of TbFeO$_{3}$ as a function of magnetic field between 5 and 10 K and up to 1.5~T. The sweep rate of the magnetic field used here was 50~mT per minute. Measurements were taken isothermally after zero field cooling.
\end{figure}

\begin{figure}[htbp]
	\centering
  	\begin{minipage}[t]{7.2 cm}	       
      \includegraphics[width=1\textwidth]{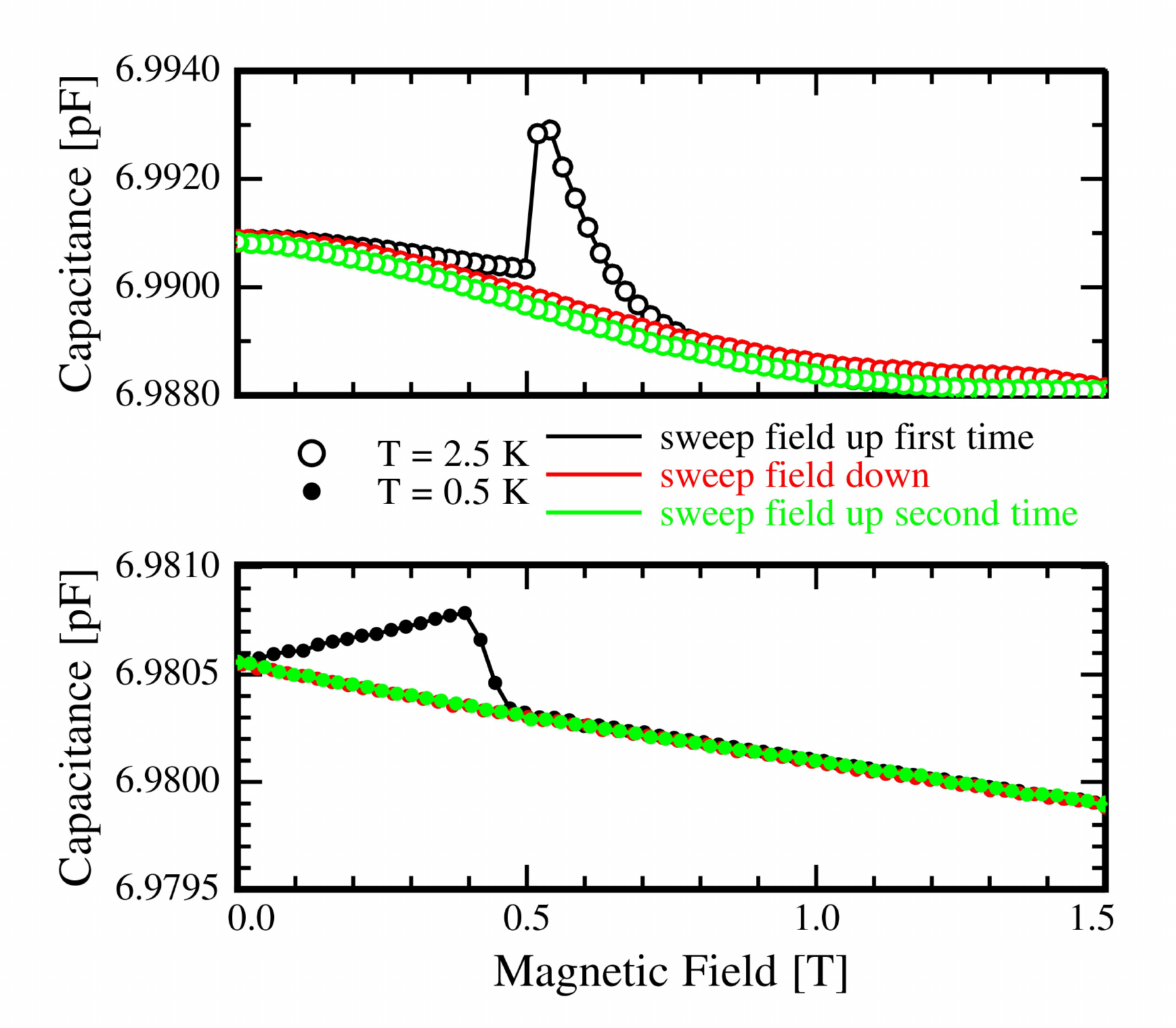}
   \end{minipage}
   \begin{minipage}[t]{7.2 cm}
     \includegraphics[width=1\textwidth]{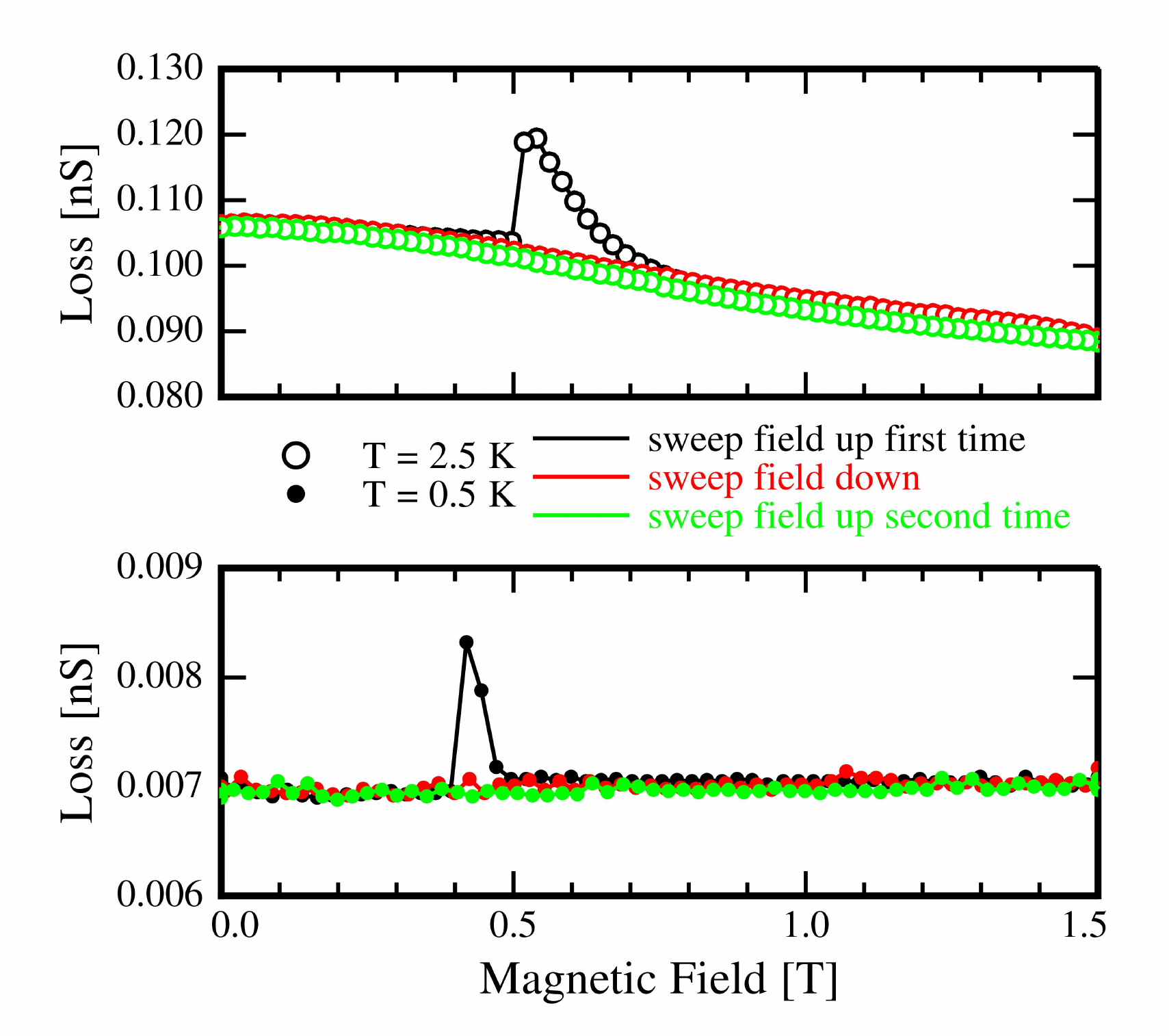}
	\end{minipage}
  \label{fig:BScanLow}
	\raggedright

FIG. S5: Example data of the capacitance and loss measured at 0.5 and 2.5~K up to 1.5~T.  A clear peak is observed for data measured after zero field cooling at 0.5~T, consistent with the transition from the commensurate to the LT-IC phases observed using neutron diffraction.  The same peak is not observed when the field was ramped down from 1.5~T and neither is it found if we ramp the field up again.  This behavior was consistent for a series of measurements between 0.3 to 2.5K and suggests that over this temperature range the LT-IC phase once entered into with field, is stable down to zero field.  The sweep rate of the magnetic field for these measurements was 50~mT per minute.
\end{figure}

\end{document}